\documentclass[a4paper,11pt]{article}
\usepackage{pos}

\newcommand{\ttbar}{\ensuremath{\mathrm{t}\overline{\mathrm{t}}}}
\newcommand{\ttH}{\ensuremath{\ttbar \mathrm{H}}}
\newcommand{\tH}{\ensuremath{\mathrm{t} \mathrm{H}}}
\newcommand{\fbinv}{\ensuremath{\mathrm{fb^{-1}}}}

\title{\ttH/\tH\ production at CMS}

\author*[1]{Angela Giraldi}

\affiliation{Deutsches Elektronen-Synchrotron (DESY),\\
  Notkestrasse 85, Hamburg, Germany}
\note{For the CMS Collaboration.}

\emailAdd{angela.giraldi@cern.ch}

\abstract{
After the Higgs boson discovery, a main focus at the CERN LHC has been the measurement of its properties. Observing the Higgs boson associated production with a top quark-antiquark pair (\ttH) is particularly interesting because it provides tree-level access to measuring the Higgs boson--top quark Yukawa coupling. However, the production of a single top quark in association with the Higgs boson (\tH) is also sensitive to the sign of this coupling. This coupling is essential for the stability of the Higgs potential at high energy scales and can also be a probe for physics beyond the standard model (SM). The latest measurements of the \ttH\ and \tH\ associated production rates performed by the CMS Collaboration with proton--proton collision events at $\sqrt{s} =13$ TeV in the diphoton and multilepton channels are presented here. Events are categorized according to the lepton and jet multiplicities, and multivariate classifiers are used to distinguish signal from background processes.
The results are consistent with the SM predictions and feature the first observation of \ttH\ production in a single Higgs boson decay channel along with the first tests of its CP structure.
}

\FullConference{%
  The Tenth Annual Conference on Large Hadron Collider Physics - LHCP2022\\
  16-20 May 2022\\
  online
}

\begin{document}

\renewcommand{\logo}{\relax}
\maketitle

\section{Introduction}

With the discovery of the Higgs boson (H) by the ATLAS~\cite{bib:ATLAS,bib:ATLASHiggs} and CMS~\cite{bib:CMS,bib:CMSHiggs,CMS:2013btf} Collaborations in 2012, the last missing piece of the standard model (SM) of particle physics was observed. 
The thorough investigation of its properties is still of crucial relevance. These proceedings focus on the Yukawa interaction ($y_t$) between the Higgs boson and the top quark. 
The production of the Higgs boson in association with a top quark--antiquark pair (\ttH) or with a single top quark (\tH) involve Higgs-boson -- top-quark couplings at tree level.
In the SM, the \tH\ production cross section is expected to be particularly small, due to the negative interference of the two leading order diagrams, one in which the Higgs boson couples to a top quark, and one in which it couples to a W boson. 
However, if the relative sign of these couplings is inverted, the interference becomes constructive, hence boosting the cross section by an order of magnitude. 
The measurement of \tH\ production thus offers sensitivity to the sign of the top Yukawa coupling $y_t$.

The \ttH\ process has a complex and rich variety of final states resulting from the combinations of \ttbar\ and Higgs boson decay channels. The first observation of \ttH\ was  based on a combined analysis of proton-proton collision data at $\sqrt{s} =7,8$ and $13\,$TeV recorded with the CMS detector and corresponding to an integrated luminosity of up to 5.1, 19.7, and 35.9 \fbinv, by taking into account the Higgs boson decays into pairs of W and Z bosons, photons, $\tau$ leptons, or bottom quark jets \cite{bib:CMSttHdiscovery}. Since then, the measurements in the $\gamma\gamma$ and multilepton final states of the Higgs boson have been updated using the full LHC Run-2 data set, which corresponds to an integrated luminosity of \mbox{138\,\fbinv.} These proceedings provide an overview of the latest results.

\section{Measurement in the \ttH\,($\mathrm{H}\to \gamma\gamma$) final state}
\begin{figure}
    \centering
    \includegraphics[width=0.32\textwidth]{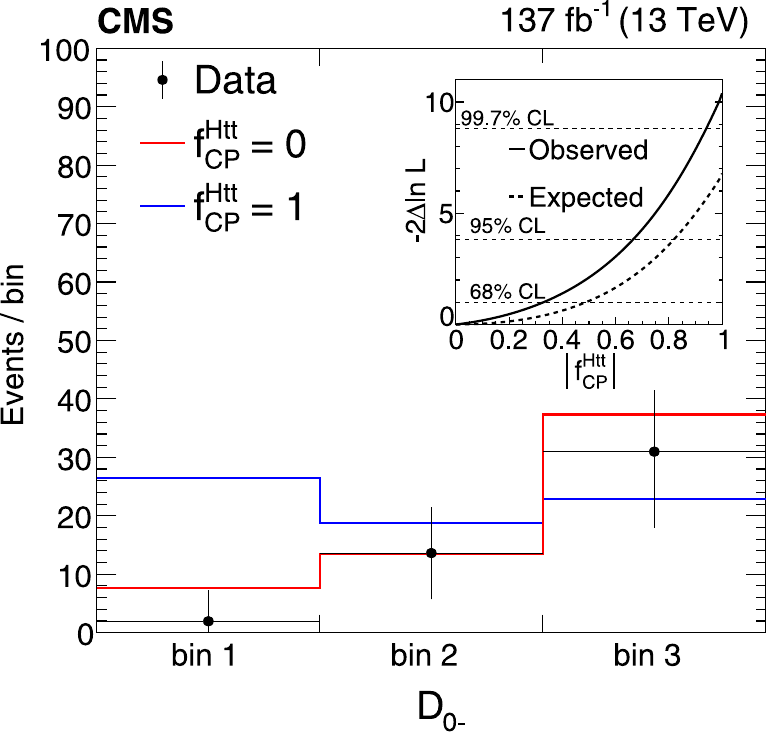}
    \qquad \qquad
    \includegraphics[width=0.45\textwidth]{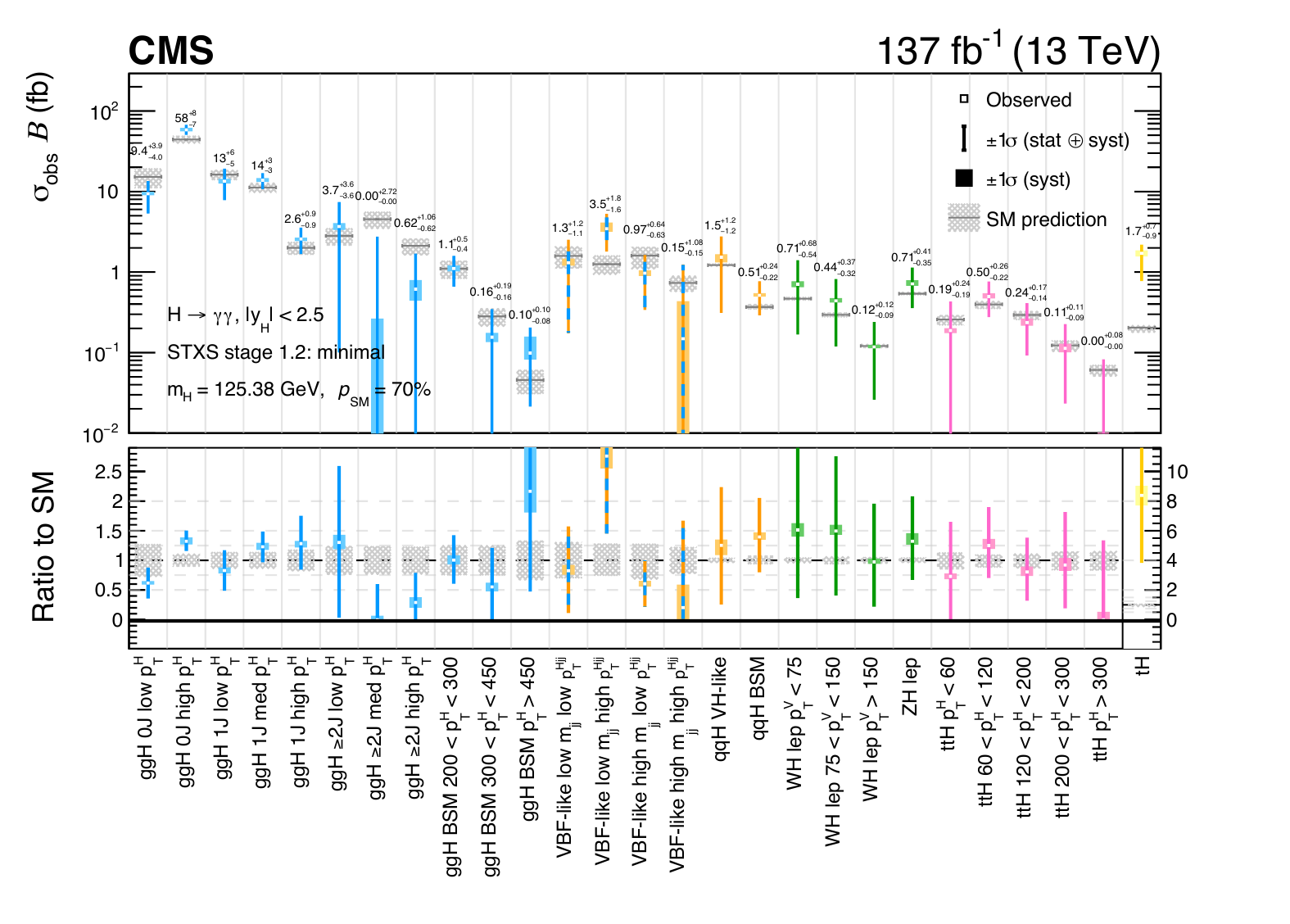}
    \caption{Left: event distributions in bins of the $\mathcal{D}_{0-}$ discriminant; the inner panel shows the likelihood scan for $f_{\mathrm{CP}}^{\mathrm{Htt}}$ \cite{bib:ttHdiphoton}. Right: observed results of the STXS fit; the lower panel shows the ratio of the fitted values to the SM predictions \cite{bib:hgammagammaSTXS}.}
    \label{fig:Do}
\end{figure}

Despite the low branching fraction of the Higgs boson decaying into photons, diphoton signatures can be measured very precisely with the CMS detector, thus allowing to resolve the Higgs boson system in \ttH($\mathrm{H}\to \gamma\gamma$) candidate events with little ambiguity. A measurement of \ttH\ production~\cite{bib:ttHdiphoton} is performed using events in which the diphoton invariant mass is required to be in the range of \mbox{100 < $m_{\gamma\gamma}$ < 180 GeV}, and the top quark pair final state includes the fully hadronic and lepton+jets channels. A boosted decision tree (BDT) is trained in each channel to discriminate between signal and background events. Only events with high BDT scores are further classified into four regions in the lepton+jets and hadronic final states, respectively. A simultaneous maximum likelihood fit based on the diphoton invariant mass distributions is performed using a double-sided Crystal Ball parametrization, to extract the signal strength $\mu_{\ttH}$, defined as the ratio of the measured and the expected cross section. This yields a signal strength $\mu_{\ttH} = 1.38 ^{+0.36}_{-0.29}$, corresponding to an observed (expected) significance of 6.6 (4.7) standard deviations, marking the first observation of the \ttH\ process in a single Higgs boson decay channel.
The CP structure of the Higgs boson couplings to fermions is also tested by using a dedicated BDT trained to distinguish between CP-even and CP-odd contributions to the coupling.
The sample is categorized in bins of the BDT output representing the $\mathcal{D}_{0-}$ observable~\cite{PhysRevD.94.055023}, and a simultaneous fit to the diphoton mass distributions is performed with $\mu_{\ttH}$ left unconstrained.
The results are shown in  Figure \ref{fig:Do} (left), with the fractional contribution of the CP-odd component measured to be  $f_{\mathrm{CP}}^{\mathrm{Htt}}=0.00 \pm 0.33$, with $|f_{\mathrm{CP}}^{\mathrm{Htt}}|<0.67$ at \mbox{95\% confidence level (C.L.).} The hypothesis of a pure pseudoscalar CP-odd structure of the Htt coupling  ($f_{\mathrm{CP}}^{\mathrm{Htt}}=1$) is excluded at 3.2 standard deviations.

In the context of measurements of Higgs boson production cross sections and couplings in events where the Higgs boson decays into a pair of photons, the first measurement of the \ttH\ production in five regions of the Higgs boson transverse momentum is performed \cite{bib:hgammagammaSTXS}. The same categorization as in Ref.\,\cite{bib:ttHdiphoton} is used to construct \ttH\ enriched categories, which are further divided to provide sensitivity to individual simplified-template-cross-section (STXS) bins. The \tH\ process is also measured in a dedicated analysis category enriched in tHq events where the top decays leptonically. The results are shown in Figure \ref{fig:Do} (right): the observed (expected) upper limit for the \tH\ production cross section at 95\% C.L. is found to be 14 (8) times the SM prediction.

\section{Measurements in the multilepton final state}
The search for \ttH\ production in multilepton final states makes use of Higgs boson decays into WW$^*$, ZZ$^*$, and $\tau^+\tau^-$, and semileptonic or hadronic decays of the top quark pair \cite{bib:multilepton}. 
Although the signal has a significantly larger branching ratio than $\mathrm{H}\to \gamma\gamma$ decay, this analysis suffers from considerable background contributions.
Events are selected and classified into ten disjoint signal regions, according to the lepton (e and $\mu$) and hadronically decaying $\tau$ ($\tau_h$) multiplicities; among these, three experimental categories are constructed to increase the sensitivity to \tH\ events. 
In the \tH\ enriched categories, artificial neural networks are used to separate signal from background, and 
then events are further categorized based on the lepton flavor and the b-tagged jet multiplicity. The rest of the categories uses a BDT that discriminates the \ttH\ signal process from background processes.
The reducible backgrounds of the analysis largely consists of contributions from non-prompt leptons and misidentified $\tau_h$ leptons, as well as electrons of misidentified charge. The contribution from these processes in the signal regions is estimated using a data-driven method. Irreducible backgrounds are dominated by \ttbar W and \ttbar Z production in the main signal regions, and dedicated control regions are used to constrain their contributions.

\begin{figure}[!b]
    \centering
    \includegraphics[width=0.32\textwidth]{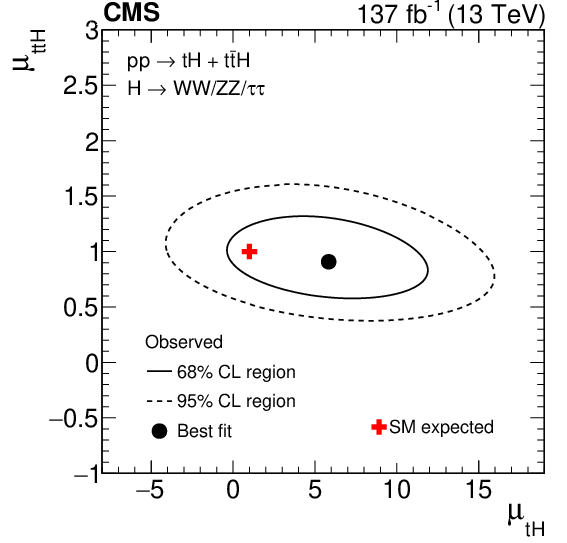}
    \qquad \qquad
    \includegraphics[width=0.39\textwidth]{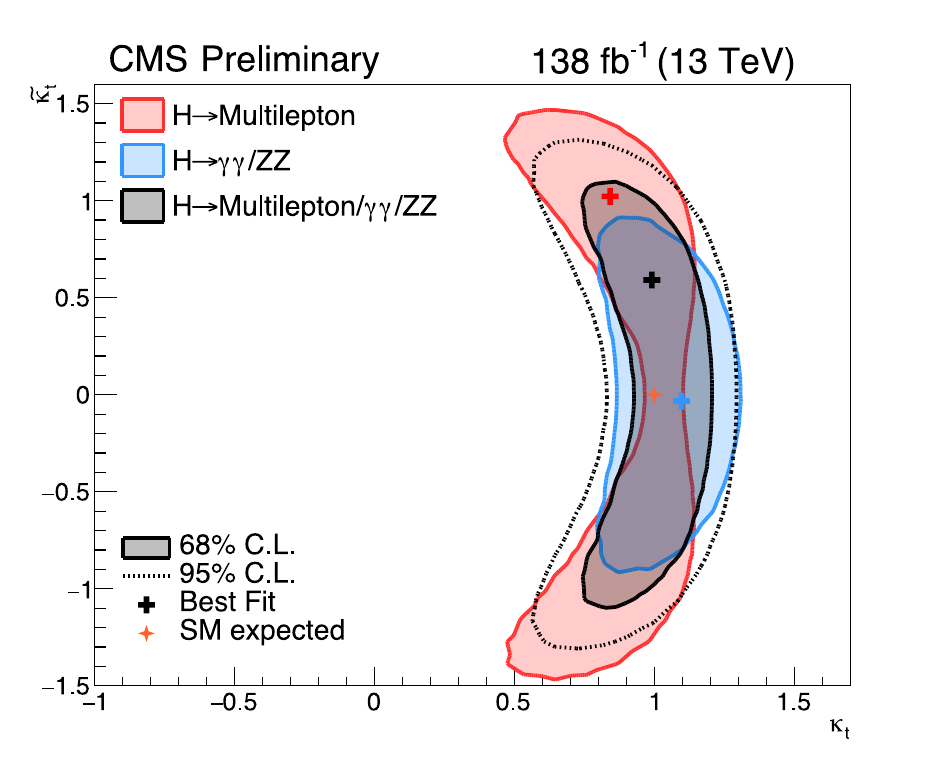}
    \caption{Left: Two-dimensional contours of the likelihood function as a function of the production rates of the \ttH\ and \tH\ signal processes \cite{bib:multilepton}. Right: Likelihood scan as function of $\kappa_t$ and $\tilde{\kappa}_t$ and two-dimensional confidence intervals for single final states and the combination of the three channels \cite{bib:CPmultilepton}.}
    \label{fig:multilepton}
\end{figure}

The signal strengths are extracted by performing a maximum likelihood fit to the data in all signal and control region subcategories.
The fitted signal strength is 0.92 $\pm$ 0.19 (stat) $^{+0.17}_{-0.13}$~(syst) for the \ttH\ signal and 5.7 $\pm$ 2.7 (stat) $\pm$ 3.0 (syst) for the \tH\ signal, and is thus compatible compatible with the SM expectations (see Figure \ref{fig:multilepton} (left)).
The values for the parameters modifying  the \ttbar W and \ttbar Z production rates after the fit to the data are $\mu_{\ttbar W}$ = 1.43 $\pm$ 0.21 (stat+syst) and \mbox{$\mu_{\ttbar Z}$ =
1.03 $\pm$ 0.14 (stat+syst).}
The observed (expected) significance for \ttH\ production is 4.7 (5.2) standard deviations, while it is 1.4 (0.3) for the \tH\ signal. The observed event yields are also interpreted in terms of coupling modifiers of the Higgs boson to the top quark ($\kappa_t$) and the W and Z bosons ($\kappa_V$). The compatibility of the observed data is compared to a grid of $(\kappa_t, \kappa_V)$ hypotheses, computing a likelihood score to construct C.L. regions in the $(\kappa_t, \kappa_V)$ plane and confidence intervals in $\kappa_t$. The measurements  constrains  \mbox{$-$0.9 < $\kappa_t$ < $-$0.7} or \mbox{ 0.7 < $\kappa_t$ < 1.1} at 95\% C.L. 

Additionally, a measurement of the CP structure of the Yukawa interaction between the Higgs  boson and one or two top quarks is performed \cite{bib:CPmultilepton}.
Machine learning techniques are applied to enhance the separation of CP-even from CP-odd scenarios. Two-dimensional confidence regions  are set on the ratios $\kappa_t$ and $\tilde{\kappa}_t$ of the couplings 
of CP-even and CP-odd Lagrangian terms, respectively, to the SM expectation for
the top-Higgs Yukawa coupling. These regions are found to be \mbox{$-$1.09 < $\kappa_t$ < $-$0.74} or \mbox{0.77 < $\kappa_t$ < 1.30,} \mbox{$-$1.4 < $\tilde{\kappa}_t$ < 1.4} at 95\% C.L. 
Similar to the $\mathrm{H}\to \gamma\gamma$ analysis, the fractional contribution of pseudo-scalar couplings is measured to be $|f_{\mathrm{CP}}^{\mathrm{Htt}}|=0.59$, with an interval of (0.24, 0.81) at 68\% C.L. 
The results are combined with previously published analyses of the $H\rightarrow \gamma\gamma$ \cite{bib:ttHdiphoton} and $H\rightarrow ZZ$ \cite{bib:hzz} decay modes, yielding two- and one-dimensional confidence regions on $\kappa_t$ and $\tilde{\kappa}_t$, as shown in Figure \ref{fig:multilepton} (right). The modifier $|f_{\mathrm{CP}}^{\mathrm{Htt}}|$ is measured to be 0.28, with an interval of $|f_{\mathrm{CP}}^{\mathrm{Htt}}|<0.55$
at 68\% C.L.

\section{Conclusion}
The Yukawa interactions between the top quark and Higgs boson are crucial to establish the validity of the standard model (SM) of particle physics. This contribution report the measurements of the production of the Higgs boson in association with a top quark--antiquark pair (\ttH) or with a single top quark (\tH), in the diphoton and multilepton final states, using the data set collected by the CMS experiment during the second data-taking period and with improved analysis techniques. The results obtained are consistent with the SM expectations.

\bibliographystyle{JHEP}
\bibliography{main}

\end{document}